\documentclass[twocolumn, pra, showpacs,superscriptaddress]{revtex4}
\usepackage{graphicx}
\usepackage{dcolumn}
\usepackage{bm}
\usepackage{amsmath}

\begin{document}

\title{Yang-Yang thermodynamics of a Bose--Fermi Mixture}
\author{Xiangguo Yin}
\affiliation{Institute of Theoretical Physics, Shanxi University, Taiyuan 030006, P. R.
China}
\author{Shu Chen}
\affiliation{Institute of Physics, Chinese Academy of Sciences, Beijing 100190, P. R.
China}
\author{Yunbo Zhang}
\email{ybzhang@sxu.edu.cn}
\affiliation{Institute of Theoretical Physics, Shanxi University, Taiyuan 030006, P. R.
China}

\begin{abstract}
We investigate theoretically the behavior of a one-dimensional interacting
Bose-Fermi mixture with equal masses and equal repulsive interactions
between atoms at finite temperature in the scheme of thermodynamic Bethe
Ansatz. Combining the Yang-Yang thermodynamic formalism with local density
approximation in a harmonic trap, we calculate the density distribution of
bosons and fermions numerically by treating the radially and axially excited
states as discrete and continuous ones, respectively. Our result from
exactly solvable solutions may be used as a touchstone of 1D interacting
Bose-Fermi mixture for experimental data fitting where mean-field
theoretical approaches fail.
\end{abstract}

\maketitle


\section{Introduction}

Recent years have witnessed great development of laser cooling and optical
trapping technology, with a remarkable achievement being the quasi
one-dimensional (1D) quantum gases realized by tightly confining the
particle motion in the other two directions to zero point oscillation \cite%
{Gorlitz,Moritz,Paredes,Kinoshita,Tolra,Olshanii}. Meanwhile the Feshbach
resonance may be used to manipulate the interparticle scattering length by
simply tuning an external magnetic field, which enables us to explore the
cold atomic gases in the full interaction regime from weakly to strongly
interacting limit. In particular, a quasi-1D quantum gas of strongly
interacting bosons has been observed in the so-called Tonks-Girardeau (TG)
regime \cite{Paredes,Kinoshita}. On the other hand, mixture of quantum
degenerate gases form novel quantum many-body systems with rich phase
structures. Particularly interesting systems are the Bose-Fermi mixtures
\cite{Molmer,Ospelkaus}, which rarely occur in nature and have become
experimentally accessible with the development of sympathetic cooling \cite%
{Jin,Truscott}. Theoretical investigations on the quasi-1D Bose-Fermi
mixture have focused on their phase diagrams and ground-state properties in
the scheme of Luttinger liquid theory \cite{Cazalilla,Mathey} and Bethe
ansatz method \cite{Imambekov,Guan,Frahm}.

The ground state energy of a 1D system of bosons with repulsive
delta-function interaction was first calculated by Lieb and Lininger \cite%
{Lieb}. Yang and Yang extended this Bethe ansatz method to finite
temperature 40 years ago \cite{YangYang}. The Yang-Yang thermodynamic
formalism, also known as the thermodynamic Bethe ansatz (TBA), allows to
evaluate the thermodynamic properties of the 1D system. It has triggered
numerous further investigations and has been generalized later to spin-1/2
fermions \cite{Lai1}, mixture of spin-1/2 fermions and bosons \cite{Lai2},
and two-component bosons \cite{Gu}, etc. The first direct comparison between
experiments and theory based on the Yang-Yang exact solutions was carried
out last year \cite{Amerongen} in the weakly interacting Bose gas on an atom
chip and for a wide parameter range where conventional models fail to
quantitatively describe in situ measured spatial density profiles. In view
of the experimental progress, it is thus quite desirable to study
theoretically the thermodynamic properties of the quasi-1D Bose-Fermi
mixture, which might provide theoretical guidance on the potential
experimental implementation. From pure theoretical point of view, it is also
very interesting to study how robust of the Bose-Fermi phase separation,
which was predicted to appear in the limit of zero temperature \cite%
{Imambekov}, against the temperature effect.

In this paper, we combine the TBA and local density approximation (LDA) to
study the density distribution of a mixture of bosons and polarized fermions
trapped in a harmonic trap at finite temperature. Here we have the
temperature, particle number of bosons and fermions, and the interaction
strength as variables. We find that the observation of Bose-Fermi phase
separation requires even lower temperature attainable by present cooling
techniques.

The article is organized as follows. In Section II, we rederive the
thermodynamic Bethe ansatz solution by means of Yang-Yang thermodynamic
formalism for Bose-Fermi mixture where both bosons and fermions are spin
polarized. Section III describes our numerical procedure on how to
numerically evaluate the density distribution of bosons and fermions in a
harmonic trap at finite temperature. We treat in a different way for
radially and axially excited states. Finally in Section IV, we analyze the
low temperature behavior of the 1D Bose-Fermi mixture for realistic
experimental situation and make concluding remarks in Section V.

\section{Thermodynamic Bethe Ansatz of Bose-Fermi mixture}

We study a 1D interacting Bose-Fermi mixture on a line of length $L$ with
periodic boundary conditions, described by the Hamiltonian
\begin{eqnarray}
H &=&\int_{0}^{L}dx\left\{ \frac{\hbar ^{2}}{2m_{b}}\partial _{x}\Psi
_{b}^{\dag }\partial _{x}\Psi _{b}+\frac{\hbar ^{2}}{2m_{f}}\partial
_{x}\Psi _{f}^{\dag }\partial _{x}\Psi _{f}\right.  \notag \\
&&\left. +\frac{1}{2}g_{bb}\Psi _{b}^{\dag }\Psi _{b}^{\dag }\Psi _{b}\Psi
_{b}+g_{bf}\Psi _{b}^{\dag }\Psi _{f}^{\dag }\Psi _{f}\Psi _{b}\right\} ,
\label{H2}
\end{eqnarray}%
where $\Psi _{b}$, $\Psi _{f}$ are field operators for a boson of mass $%
m_{b} $ and for a fermion of mass $m_{f}$, and $g_{bb}$, $g_{bf}$ are
boson-boson and boson-fermion interaction strengths, respectively. Fermions
are spin-polarized so that Pauli principle excludes their $s$-wave
interaction ($g_{ff}=0$). This model is exactly solvable for equal masses
and equal repulsive boson--boson and boson--fermion interaction strengths,
i.e.
\begin{equation}
m_{b}=m_{f}=m,\text{\qquad }g_{bb}=g_{bf}=g.  \label{condition}
\end{equation}%
Although an exact solution is available only under conditions (\ref%
{condition}), deviations slightly from this integrable line are expected not
to dramatically change the characteristic properties of the system, e.g.,
the phase separation. Following Lai-Yang's original convention, we assume $%
2m=1$ and $\hbar =1$ and write the Hamiltonian (\ref{H2}) in its first
quantization form%
\begin{equation}
H=-\sum_{i=1}^{N}\frac{\partial ^{2}}{\partial x_{i}^{2}}+2c\sum_{i<j}\delta
\left( x_{i}-x_{j}\right) ,  \label{H1}
\end{equation}%
with $c=mg/\hbar ^{2}$. Among the $N$ particles there are $M$ bosons and the
rest of them are fermions. The many-body wave function is supposed to be
symmetric under odd permutations with respect to indices $i=\left\{
1,...,M\right\} $ (bosons) and antisymmetric with respect to $i=\left\{
M+1,...,N\right\} $ (fermions).

For periodic boundary conditions, Imambekov and Demler obtained the
following set of Bethe Ansatz equations (BAE) \cite{Imambekov}%
\begin{equation}
\exp \left( ik_{j}L\right) =\prod\limits_{\beta =1}^{M}\frac{k_{j}-\Lambda
_{\beta }+ic/2}{k_{j}-\Lambda _{\beta }-ic/2},\qquad j=1,...,N,  \label{BA1}
\end{equation}%
\begin{equation}
1=\prod\limits_{i=1}^{N}\frac{k_{i}-\Lambda _{\alpha }+ic/2}{k_{i}-\Lambda
_{\alpha }-ic/2},\qquad \alpha =1,...,M,  \label{BA2}
\end{equation}%
where the momenta $k_{1},...k_{N}$ are a set of unequal numbers, and
spectral parameters $\Lambda _{1},...\Lambda _{M}$ are the analogs of the
momenta. It has also been proved that all solutions of (\ref{BA1}-\ref{BA2})
are always real, which simplifies greatly the analysis at finite temperature.

For thermodynamics at finite temperature we use the BAE to derive a set of
nonlinear integral equations, i.e. TBA equations, which describes the
thermodynamics of the model at finite temperature. Taking logarithm of the
BAE (\ref{BA1}-\ref{BA2}), we arrive at the following discrete Bethe ansatz
equations \
\begin{align}
k_{j}L& =2\pi I_{j}+\sum_{\beta =1}^{M}\theta \left( 2k_{j}-2\Lambda _{\beta
}\right) ,  \notag \\
2\pi J_{\alpha }& =\sum_{i=1}^{N}\theta \left( 2k_{i}-2\Lambda _{\alpha
}\right) ,  \label{BA eq}
\end{align}%
with $\theta \left( k\right) =-2\arctan \left( k/c\right) $. Here $I_{j}$
and $J_{\alpha }$ are integer or half integer quantum numbers (depending on
the parity of $M$ and $N$), which play the role of quantum numbers for the
momentum $k$ and spectral parameter $\Lambda $ respectively. For a
particular configuration, if an arbitrary quantum number is chosen, it is
either occupied (in the set of quantum numbers for the system), called a
root, or not occupied, called a hole. In the thermodynamic limit, the
distributions of momentum and spectral parameter become dense, and it is
convenient to introduce the density functions of roots and holes,
respectively. We denote with $\rho \left( k\right) $ and $\rho _{h}\left(
k\right) $ the density functions of the momentum $k$ and holes, and with $%
\sigma \left( \Lambda \right) $ and $\sigma _{h}\left( \Lambda \right) $ the
density functions of spectral parameter $\Lambda $ and its holes. They are
defined by
\begin{align*}
L\left( \rho \left( k\right) +\rho _{h}\left( k\right) \right) dk& =dI, \\
L\left( \sigma \left( \Lambda \right) +\sigma _{h}\left( \Lambda \right)
\right) d\Lambda & =dJ.
\end{align*}%
Differentiate equation (\ref{BA eq}) with respect to $k$ and $\Lambda $\
separately, we obtain a set of coupled integral equations%
\begin{align}
\rho \left( k\right) +\rho _{h}\left( k\right) & =\frac{1}{2\pi }+\frac{1}{%
2\pi }\int_{-\infty }^{\infty }K\left( k,\Lambda \right) \sigma \left(
\Lambda \right) d\Lambda ,  \notag \\
\sigma \left( \Lambda \right) +\sigma _{h}\left( \Lambda \right) & =\frac{1}{%
2\pi }\int_{-\infty }^{\infty }K\left( \Lambda ,k\right) \rho \left(
k\right) dk,  \label{constraint}
\end{align}%
where%
\begin{equation*}
K\left( x,y\right) =\frac{4c}{c^{2}+4\left( x-y\right) ^{2}}.
\end{equation*}%
The total number of particles and that of bosons per unit length can be
obtained by integrating the density functions of momentum and spectral
parameter as follows%
\begin{equation}
N/L=\int_{-\infty }^{\infty }\rho \left( k\right) dk,\qquad
M/L=\int_{-\infty }^{\infty }\sigma \left( \Lambda \right) d\Lambda ,
\label{particle number}
\end{equation}%
while the energy of the system per unit length is given by%
\begin{equation*}
E/L=\int_{-\infty }^{\infty }k^{2}\rho \left( k\right) dk.
\end{equation*}%
With the help of the approach first introduced by Yang and Yang, the entropy
of the present model at finite temperature is%
\begin{align*}
S/L& =\int \left[ \left( \rho +\rho _{h}\right) \ln \left( \rho +\rho
_{h}\right) -\rho \ln \rho -\rho _{h}\ln \rho _{h}\right] dk \\
& +\int \left[ \left( \sigma +\sigma _{h}\right) \ln \left( \sigma +\sigma
_{h}\right) -\sigma \ln \sigma -\sigma _{h}\ln \sigma _{h}\right] d\Lambda .
\end{align*}%
The Gibbs free energy of the model is then defined by $F=E-TS-\mu _{F}\left(
N-M\right) -\mu _{B}M$, where $\mu _{F}$ and $\mu _{B}$ are two Lagrange
multipliers, and $T$ is the temperature. In order to arrive the thermal
equilibrium, we minimize the free energy with respect to the density
functions $\rho \left( k\right) $ and $\sigma \left( \Lambda \right) $
subject to the constraint (\ref{constraint}). In addition, the numbers of
fermions and bosons are kept to be constants respectively. It can be proved
rigorously that $\mu _{F}$ and $\mu _{B}$ are the chemical potentials of
fermions and bosons, respectively.

Applying the minimum condition $\delta F=0$ gives rise to the following
nonlinear integral equations, i.e. TBA equations%
\begin{align}
\epsilon \left( k\right) & =-\mu _{F}+k^{2}-\frac{T}{2\pi }  \notag \\
& \times \int_{-\infty }^{\infty }K\left( k,\Lambda \right) \ln \left(
1+\exp \left( -\varphi \left( \Lambda \right) /T\right) \right) d\Lambda ,
\notag \\
\varphi \left( \Lambda \right) & =\mu _{F}-\mu _{B}-\frac{T}{2\pi }  \notag
\\
& \times \int_{-\infty }^{\infty }K\left( \Lambda ,k\right) \ln \left(
1+\exp \left( -\epsilon \left( k\right) /T\right) \right) dk,
\label{TBA_nonlinear}
\end{align}%
where we have defined%
\begin{eqnarray}
\exp \left( \epsilon \left( k\right) /T\right)  &=&\rho _{h}\left( k\right)
/\rho \left( k\right) ,\text{ }  \notag \\
\exp \left( \varphi \left( \Lambda \right) /T\right)  &=&\sigma _{h}\left(
\Lambda \right) /\sigma \left( \Lambda \right) ,  \label{epsilon}
\end{eqnarray}%
and the set of equations (\ref{constraint}) becomes%
\begin{align}
2\pi \rho \left( k\right) \left( 1+\exp \left( \epsilon \left( k\right)
/T\right) \right) & =1+\int_{-\infty }^{\infty }K\left( k,\Lambda \right)
\sigma \left( \Lambda \right) d\Lambda ,  \notag \\
2\pi \sigma \left( \Lambda \right) \left( 1+\exp \left( \varphi \left(
\Lambda \right) /T\right) \right) & =\int_{-\infty }^{\infty }K\left(
\Lambda ,k\right) \rho \left( k\right) dk.  \label{BA_linear}
\end{align}%
The density functions $\rho \left( k\right) $ and $\sigma \left( \Lambda
\right) $ can be obtained by solving the above coupled integral equations (%
\ref{TBA_nonlinear}) and (\ref{BA_linear}). The TBA approach described here
is universal for discussing the thermodynamics of 1D integrable model. Once $%
T,c,\mu _{F}$ and $\mu _{B}$ are determined, all thermodynamic properties
are known. For instance, the pressure and the free energy are
\begin{align}
P& =\frac{T}{2\pi }\int_{-\infty }^{\infty }\ln \left( 1+\exp \left( -\frac{%
\epsilon \left( k\right) }{T}\right) \right) dk,  \notag \\
F& =-PL+M\mu _{B}+\left( N-M\right) \mu _{F}.  \label{energy}
\end{align}

\section{Local density approximation and numerical procedure}

Starting from the solution to the TBA equations derived above for a uniform
1D Bose-Fermi mixture in thermal equilibrium, in this section we aim to find
the numerical results for the finite-temperature density distributions of
bosons $n_{B}\left( x\right) $ and fermions $n_{F}\left( x\right) $ under
the local density approximation (LDA).

Firstly we describe our iteration process in solving the TBA integral
equations. Due to the fact that the TBA equations are a set of coupled
nonlinear integral equations there exist no closed analytical solutions for
them. Nonetheless, numerically this is in principle a well controllable
problem (it becomes, however, quite complex for an increasing number of
particle types) and we here solve the equations by iteration. The
convergence of this procedure to a solution and its very existence have been
investigated most naturally by means of the Banach fixed point theorem \cite%
{Fring}.

The iteration process is as follows. For given $T,c,\mu _{F}$ and $\mu _{B}$%
, we initialize $\epsilon \left( k\right) $ and $\varphi \left( \Lambda
\right) $ on the rhs of equations (\ref{TBA_nonlinear}) with the
corresponding zero-temperature trial functions $\epsilon ^{\left( 0\right)
}\left( k\right) =-\mu _{F}+k^{2}$ and $\varphi ^{\left( 0\right) }\left(
\Lambda \right) =\mu _{F}-\mu _{B}$, respectively. In a first step, we
obtain $\epsilon ^{\left( 1\right) }\left( k\right) $ and $\varphi ^{\left(
1\right) }\left( \Lambda \right) $ on the lhs of equations (\ref%
{TBA_nonlinear}) and let them be the new trial functions. The scheme
continues with updates to $\epsilon ^{\left( n\right) }\left( k\right) $ and
$\varphi ^{\left( n\right) }\left( \Lambda \right) $ with $n=1,2,...$. When
the relative error between $\epsilon ^{\left( n\right) }\left( k\right) $
and $\epsilon ^{\left( n+1\right) }\left( k\right) $ and that between $%
\varphi ^{\left( n\right) }\left( \Lambda \right) $ and $\varphi ^{\left(
n+1\right) }\left( \Lambda \right) $ reach a small quantity, e.g., $10^{-20}$%
, sufficient convergence is obtained and $\epsilon ^{\left( n\right) }\left(
k\right) $ and $\varphi ^{\left( n\right) }\left( \Lambda \right) $ are
considered as the solutions of TBA equations. We then put these solutions
into equations (\ref{BA_linear}), and meanwhile the initial trial density
functions $\rho \left( k\right) $ and $\sigma \left( \Lambda \right) $ is
set as $\rho ^{\left( 0\right) }\left( k\right) =1/2\pi \left( 1+\exp \left(
\epsilon \left( k\right) /T\right) \right) $ and $\sigma ^{\left( 0\right)
}\left( \Lambda \right) =0,$ respectively. With the same iteration process,
we can obtain the solution of equations (\ref{BA_linear}), the integration
of which gives the particle density. Note that, to insure the accurateness
of the integration in the equations (\ref{TBA_nonlinear}) and (\ref%
{BA_linear}) for very small interaction strength $c$, the integrand should
be divided into more parts with interpolation method firstly and then
integrated numerically.

Experimentally 1D gases are usually achieved by trapping the atoms in a
tight harmonic trap with strong transverse confinement and weak confinement
along the axis, $\omega _{\bot }\gg \omega _{//}$. The main parameters from
the experiment \cite{Amerongen} on Yang-Yang thermodynamics of $^{87}$Rb
atoms in the $\left\vert F=2,m_{F}=2\right\rangle $ hyperfine state are
adopted here. The Bose-Fermi mixture is trapped in a harmonic potential with
$\omega _{\bot }/2\pi =3280Hz$, $\omega _{//}/2\pi =8.5Hz$.

The numerical procedure is as follows: The total linear density relies on
two contributions, i.e. that from the radial ground state and that from the
radially excited states. The density of atoms populated in the radial ground
state is obtained by solving the TBA equation, while the atoms in the
radially excited states are treated \textit{discretely} as an independent
ideal 1D Bose(Fermi) gas in thermal equilibrium. This is because our
temperature here is on the order of the radial level splitting, $\hbar
\omega _{\bot }/k_{B}=157.4nK$, so that the fraction of the atoms in
radially excited states should not be neglected. Along the axis, the energy
level splitting, $\hbar \omega _{//}/k_{B}=0.4nK$ is so small that we can
use the LDA to account for the axial potential via a \textit{continuously}
varying chemical potential $\mu \left( x\right) =\mu -V(x)$. In this way, we
obtain the total linear densities of bosons $n_{B}(x)$ and fermions $%
n_{F}(x) $ in the magnetic trap which may be used to fit the experimental
data from absorption images.

For the radial ground state, the TBA equations for uniform gas can be
applied locally to the trapped gas if the condition for the LDA are met. One
assumes that in slowly varying external harmonic trap the chemical
potentials of bosons and fermions are changed into%
\begin{eqnarray}
\mu _{B}\left( x\right) &=&\mu _{B}-\frac{1}{2}m\omega _{//}^{2}x^{2},
\notag \\
\mu _{F}\left( x\right) &=&\mu _{F}-\frac{1}{2}m\omega _{//}^{2}x^{2}.
\label{lda}
\end{eqnarray}%
Here, $\mu _{B}$ and $\mu _{F}$ are chemical potentials for bosons and
fermions in the center of the harmonic trap. Replacing the chemical
potentials in eqs. (\ref{TBA_nonlinear}) by their LDA\ values, we can obtain
numerically the density functions $\rho \left( \mu _{B},\mu _{F},x,k\right) $
and $\sigma \left( \mu _{B},\mu _{F},x,\Lambda \right) $ by iteratively
solving the TBA equations (\ref{TBA_nonlinear}) together with the constraint
(\ref{BA_linear}). The integration of these density functions yields the
axial density distributions
\begin{equation}
n_{B}^{TBA}\left( x\right) =\int_{-\infty }^{\infty }\sigma \left( \mu
_{B},\mu _{F},x,\Lambda \right) d\Lambda  \label{nbtba}
\end{equation}%
for bosons\ and
\begin{equation}
n_{F}^{TBA}\left( x\right) =\int_{-\infty }^{\infty }\rho \left( \mu
_{B},\mu _{F},x,k\right) dk-n_{B}^{TBA}\left( x\right)  \label{nftba}
\end{equation}%
for fermions, which are the LDA revisions to their uniform counterparts,
i.e. eqs. (\ref{particle number}).

Similar strategies are applied to acquire the densities of bosons and
fermions at the radially excited states. For bosons it can be expected that
the interaction will significantly affect only the distribution in the
radial ground state, while the population in the radially excited states can
be well described by the distribution of ideal Bose gas. Fermions in the
ground state do not interact with each other due to the Pauli exclusive
principle. Thus for fermions, the chemical potential $\mu _{F}$ can be even
larger than $\hbar \omega _{\bot }$, with the population in the radially
excited states described by the distribution of ideal Fermi gas. The
radially excited states are ($j+1$)-fold degenerate, i.e. for each radial
quantum number $j\geq 1$ there are $j+1$ excited states sharing the same
energy. We treat each excited state as an independent ideal 1D Bose or Fermi
gas in thermal equilibrium with the chemical potential of the gas in the
radial ground state given by
\begin{eqnarray}
\mu _{B}^{j}\left( x\right) &=&\mu _{B}\left( x\right) -j\hbar \omega _{\bot
},  \notag \\
\mu _{F}^{j}\left( x\right) &=&\mu _{F}\left( x\right) -j\hbar \omega _{\bot
},  \label{lda2}
\end{eqnarray}%
respectively. The density distributions in radially excited state ($j$) are%
\begin{align}
n_{B}^{j}\left( x\right) & =\int_{-\infty }^{\infty }\frac{1}{2\pi }\frac{1}{%
\exp \left[ \left( \frac{\hbar ^{2}}{2m}k^{2}-\mu _{B}^{j}\left( x\right)
\right) /k_{B}T\right] -1}dk,  \notag \\
n_{F}^{j}\left( x\right) & =\int_{-\infty }^{\infty }\frac{1}{2\pi }\frac{1}{%
\exp \left[ \left( \frac{\hbar ^{2}}{2m}k^{2}-\mu _{F}^{j}\left( x\right)
\right) /k_{B}T\right] +1}dk.  \label{excitedstate}
\end{align}

The total linear densities are given by summing over the TBA results for
radial ground state (\ref{nbtba}, \ref{nftba}) and the ideal-gas results for
radially excited states (\ref{excitedstate})
\begin{align}
n_{B}\left( x\right) \ & =n_{B}^{TBA}\left( x\right) +\sum_{j=1}^{\infty
}\left( j+1\right) n_{B}^{j}\left( x\right) ,  \notag \\
n_{F}\left( x\right) \ & =n_{F}^{TBA}\left( x\right) +\sum_{j=1}^{\infty
}\left( j+1\right) n_{F}^{j}\left( x\right) .  \label{density}
\end{align}

\section{Low Temperature Behavior and Experimental Considerations}

\begin{figure}[tbp]
\includegraphics[width=3.5in]{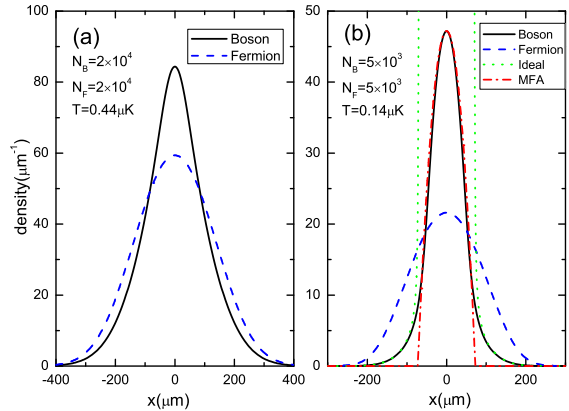}
\caption{(Color online) Bose and Fermi density distributions at high (a) and
low (b) temperatures for weakly interacting mixture. The interaction
strength is chosen as $c_{0}=0.30\protect\mu m$. Yang-Yang thermodynamics is
clearly seen for lower temperature as a narrow density peak. Also shown are
the ideal Bose-gas profile and that from a mean field theory in the
Thomas-Fermi approximation with the same peak density.}
\label{fig1}
\end{figure}

Here, as an example which is experimentally accessible, we show the linear
density of atomic clouds with 3D scattering length $a$ equal to $100a_{B}$ ($%
a_{B}$ is Bohr radius) and the mass of the atoms, both boson and fermion,
being chosen as that of $^{87}$Rb. 1D Bose--Fermi mixture have not attracted
much attention until recently, when it became possible to realize such
systems in experiments with cold atoms \cite{Ospelkaus}. Experimentalists
care more about the heteronuclear Bose-Fermi mixture, such as $^{87}$Rb-$%
^{40}$K, in which case the heteronuclear interactions can be tailored by
means of Feshbach resonances. Bose-Fermi mixture system, however, may be
composed of isotopes of atoms whose mass difference is very small, such as $%
^{6}$Li--$^{7}$Li, or $^{86}$Rb-$^{87}$Rb, etc. The exactly solvable case
considered here is relevant to current experiments, and can be used as a
benchmark to check the validity of different approaches.

We first consider the weakly interacting mixture with the effective 1D
coupling strength expressed through the 3D scattering length $a$ as $%
c=2m\omega _{\bot }a/\hbar $ if $a\ll \left( \hbar /m\omega _{\bot }\right)
^{1/2}$ \cite{Olshanii}. We denote as $c_{0}$ the coupling strength for bare
$^{87}$Rb background scattering which is approximately $0.30\mu m^{-1}$. In
Figs. 1(a) and (b), we show the linear density of atomic clouds for $N_{B}$
bosons (black solid line) and $N_{F}$ fermions (blue dashed line) in the
magnetic trap for different temperatures. Experimentally these data may be
obtained by absorption imaging and integrating the atom number along $z$%
-axis. At high temperature ($T=0.44\mu K$), population in excited states
contribute a lot to the density and the result from Yang-Yang formalism for
ground state is only a small fraction, hence is not visible. The Yang-Yang
thermodynamics is clearly seen for lower temperature ($T=0.14\mu K$) as a
narrow density peak for bosons, where both the ideal-gas and the mean field
distributions in the Thomas-Fermi approximation fail to quantitatively
describe the spatial density profiles. We also find that in the weakly
interacting limit the number of fermions does not affect the density profile
of the bosons very much. For instance, one can hardly discern the difference
in the distributions of $5\times 10^{3}$ bosons when we include in the
mixture $5\times 10^{2},5\times 10^{3}$ or $5\times 10^{4}$ fermions. This
can be attributed to the relatively small number of fermions in the ground
state.

\begin{figure}[tbp]
\includegraphics[width=3.5in]{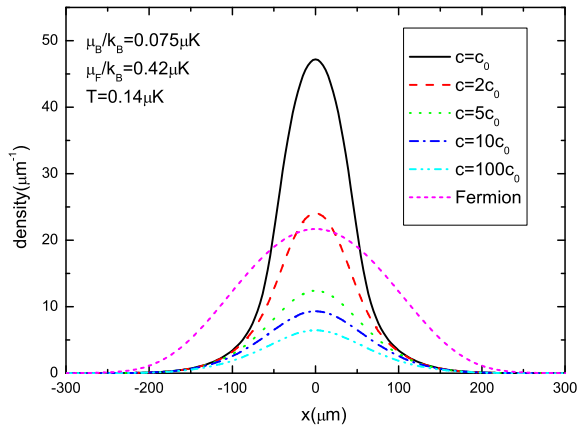}
\caption{(Color online) Bose and Fermi density distributions with fixed
chemical potential in the trap center and for different interaction
strengths. When the interaction strength increases, less number of bosons
are needed in the gas to achieve the chemical potential, while the number of
fermions in the mixture almost remain unchanged.}
\label{fig2}
\end{figure}

The interaction in the mixture, on the other hand, may be enhanced greatly
by the so-called confinement induced resonances \cite{Olshanii}.
Additionally, all interactions can be tuned using available Bose-Fermi
Feshbach resonances \cite{Ospelkaus}. In Fig. 2 we illustrate how the
bosonic density distribution will change with the increase of interaction
strength. For fixed chemical potentials of Bose and Fermi gas, as
interactions get stronger, the density of Bose gas decreases very quickly,
while the density of fermi gas keeps unchanged. This can be explained as
follows. In determining the density of the atomic gas, Pauli exclusive
principle plays a more important role than the Bose-Fermi interaction $%
g_{bf} $ for atoms in the ground state (even in the strongly interacting
limit) as a result of most fermions occupying the excited states.

Imambekov and Demler \cite{Imambekov} predicted the existence of the
Bose-Fermi phase separation at very low temperature and very strongly
interacting limit, i.e. the relative distribution of bosons and fermions
changes with interaction and the Fermi density shows strong nonmonotonous
behavior for strong interactions. They firstly get the magnon energy
spectrum for large $\gamma $( $\gamma =c/n$ and $n$ is density) and then use
the local density approximation and energy spectrum to obtain the density
distribution. There is, however, no obvious signature of this phase
separation in our Fig. 2 which is plotted for $T=0.14\mu K$ even for very
large interaction strength. Clearly the temperature smears off this
many-body quantum effect. To observe the phase separation, we need further
cool the atomic gas to even lower temperature. In Fig. 3, we compare the
density profiles for a mixture of 100 bosons and 100 fermions in the
strongly interacting limit $c=100\mu m^{-1}$ for different temperatures,
some of which are beyond the current experimental reach. One clearly sees
that the phase separation appears at $nK$ temperatures. Also shown are the
zero temperature result which is obtained analytically from Bethe ansatz
method \cite{Imambekov}.

\begin{figure}[tbp]
\includegraphics[width=3.5in]{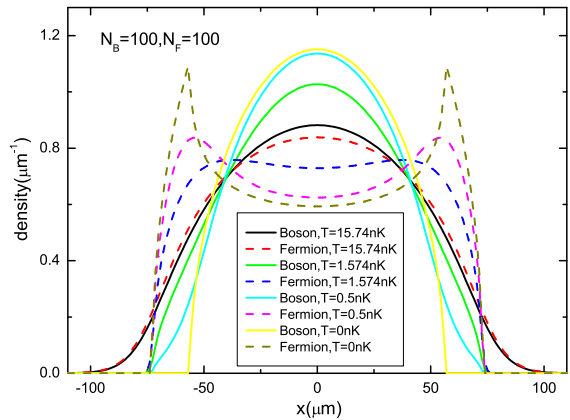}
\caption{(Color online) Bose and Fermi density distributions at different
temperatures for large interaction strength $c=100\protect\mu m^{-1}$. Phase
separation appears at very low temperature.}
\label{fig3}
\end{figure}

\begin{figure}[tbp]
\includegraphics[width=3.5in]{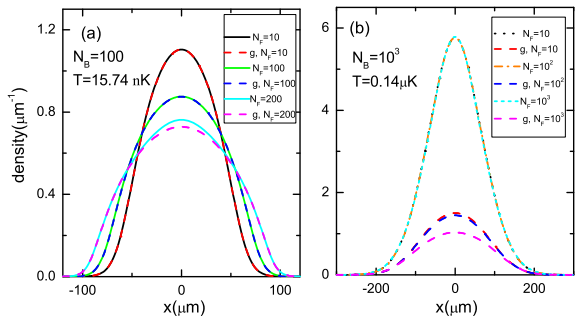}
\caption{(Color online) Modification of bosonic density by different number
of fermions in the mixture for large interaction strength $c=100 \protect\mu %
m^{-1}$. In ultra-low temperature ($nK$) few bosons would populate on the
radially excited states. For more fermions in the mixture, bosons in the
trap center move to the edge. At slightly higher temperature, the total
density remains unchanged. Here ''g" stands for density profile of atoms in
the ground state.}
\label{fig4}
\end{figure}

Finally, let us see how the number of fermions would affect the Bose density
distribution. In the weakly interacting limit, we do not observe significant
change in the density profile of the bosons when we change the number of
fermions. However, when both the strongly interacting and ultracold
conditions are met, the presence of more fermions drastically alter the
density of bosons. We show in Fig. 4(a) the modification of bosonic density
by different number of fermions. More fermions produce stronger Fermi
pressure, which tends to flatten the Bose density when the interactions
between bosons and fermions are strong. We also noticed that the ground
state density (dashed lines) coincides almost perfectly with the total
density (solid lines) for $N_{F}=10-100$, indicating that all bosons $%
N_{B}=100$ occupy the radial ground state. One can see the distinction
appears for even larger number of fermions, e.g., for $N_{F}=200$, the
fattest two curves do not match each other any more. Temperature again blurs
the quantum effect from increasing the number of fermions, i.e. the total
density remains unchanged although that of atoms in the ground state indeed
decrease for more involved fermions as can be seen in Fig. 4(b).

\section{Conclusion}

In summary, we have theoretically studied the quasi-1D system of Bose-Fermi
mixture with equal repulsion between atoms at finite temperature. For the
mixture system in a harmonic trap, we calculate the density distributions of
bosons and fermions numerically by using the combination of LDA and TBA and
treating the radially and axially excited states as discrete and continuous
ones, respectively. As the typical temperature for the ultracold gas is on
the order of the radial level splitting, the density of atoms populated in
the radial ground state is obtained by solving the TBA equation, while those
in the radially excited states are treated discretely as an independent
ideal 1D Bose gas in thermal equilibrium. The density distributions of
bosons and fermions are thus calculated in various experimental situations
and the effects due to the interaction strength, particle numbers and
temperature are discussed. Our results show that the phase separation
between bosons and fermions takes place at even lower temperature attainable
by recent cooling techniques.

\begin{acknowledgments}
This work is supported by NSF of China under Grant No. 10774095 and
10821403, NSF of Shanxi Province under grant No. 2009011002, 973 Program
under Grant No. 2006CB921102, and National Program for Basic Research of
MOST China. Y.Z. thanks Prof. Jing Zhang for helpful discussions.
\end{acknowledgments}


\begin{thebibliography}{99}
\bibitem{Gorlitz} A. G\"{o}rlitz, et.al., Phys. Rev. Lett. \textbf{87},
130402 (2001).

\bibitem{Moritz} H. Moritz, T. St\"{o}ferle, M. K\"{o}hl, and T. Esslinger,
Phys. Rev. Lett. \textbf{91}, 250402 (2003); T. St\"{o}ferle, H. Moritz, C.
Schori, M. K\"{o}hl, and T. Esslinger, ibid. \textbf{92}, 130403 (2004).

\bibitem{Paredes} B. Paredes, A. Widera, V. Murg, O. Mandel, S. F\"{o}lling,
I. Cirac, G. V. Shlyapnikov, T. W. H\"{a}nsch, and I. Bloch, Nature \textbf{%
429}, 277 (2004).

\bibitem{Kinoshita} T. Kinoshita, T. Wenger, and D. S. Weiss, Science
\textbf{305}, 1125 (2004).

\bibitem{Olshanii} M. Olshanii, Phys. Rev. Lett. \textbf{81}, 938 (1998).

\bibitem{Tolra} B. Laburthe-Tolra, K. M. O'Hara, J. H. Huckans, W. D.
Phillips, S. L. Rolston, and J. V. Porto, Phys. Rev. Lett. \textbf{92},
190401 (2004).

\bibitem{Molmer} K. Molmer, Phys. Rev. Lett. \textbf{80}, 1804 (1998) ; L.
Viverit, C. J. Pethick, and H. Smith, Phys. Rev. A \textbf{61}, 053605
(2000); M. Lewenstein, L. Santos, M. A. Baranov, and H. Fehrmann, Phys. Rev.
Lett. \textbf{92}, 050401 (2004).

\bibitem{Jin} B. DeMarco and D. S. Jin, Science \textbf{285}, 1703 (1999);
F. Schreck et al., Phys. Rev. Lett. \textbf{87}, 080403 (2001) ; G. Modugno
et al., Science \textbf{297}, 2240 (2002); Z. Hadzibabic et al., Phys. Rev.
Lett. \textbf{88}, 160401 (2002); J. Goldwin et al., Phys. Rev. A \textbf{70}%
, 021601(R) (2004).

\bibitem{Truscott} A. G. Truscott et al., Science \textbf{291}, 2570 (2001).

\bibitem{Cazalilla} M. A. Cazalilla and A. F. Ho, Phys. Rev. Lett. \textbf{91%
}, 150403 (2003).

\bibitem{Mathey} L. Mathey, D. W. Wang, W. Hofstetter, M. D. Lukin, and E.
Demler, Phys. Rev. Lett. \textbf{93}, 120404 (2004).

\bibitem{Lieb} E. H. Lieb and W. Liniger, Phys. Rev. \textbf{130}, 1605
(1963); Y. Hao, Y. Zhang, J.-Q. Liang, and S. Chen, Phys. Rev. A \textbf{73}%
, 063617 (2006); C. N. Yang, Phys. Rev. Lett. \textbf{19}, 23 (1967); M.
Gaudin, Phys. Lett. \textbf{24A}, 55 (1967); C. K. Lai and C. N. Yang, Phys.
Rev. A \textbf{3}, 393 (1971).

\bibitem{YangYang} C. N. Yang and C. P. Yang, J. Math. Phys. \textbf{10},
1115 (1969).

\bibitem{Lai1} C. K. Lai, Phys. Rev. Lett. \textbf{26,} 1472 (1971).

\bibitem{Lai2} C. K. Lai, Phys. Rev. A \textbf{8}, 2567 (1973).

\bibitem{Gu} S. J. Gu, Y. Q. Li, Z. J. Ying, and X. A. Zhao, Int. J. Mod.
Phys. B \textbf{16}, 2137 (2002).

\bibitem{Imambekov} A. Imambekov and E. Demler, Phys. Rev. A \textbf{73},
021602(R) (2006); A. Imambekov and E. Demler, Ann. Phys. \textbf{321}, 2390
(2006).

\bibitem{Guan} M. T. Batchelor, M. Bortz, X.-W. Guan, and N. Oelkers, Phys.
Rev. A \textbf{72}, 061603(R) (2005); X.-W. Guan, M. T. Batchelor, and J.-Y.
Lee, Phys. Rev. A \textbf{78}, 023621 (2008).

\bibitem{Frahm} H. Frahm and G. Palacios, Phys. Rev. A \textbf{72},
061604(R) (2005).

\bibitem{Amerongen} A. H. van Amerongen, J. J. P. van Es, P. Wicke, K. V.
Kheruntsyan, and N. J. van Druten, Phys. Rev. Lett. \textbf{100}, 090402
(2008).

\bibitem{Fring} A. Fring, C. Korff and B. J. Schulz, Nucl. Phys. B \textbf{%
549,} 579 (1999).

\bibitem{Ospelkaus} C. Ospelkaus, S. Ospelkaus, J. Phys. B: At. Mol. Opt.
Phys. \textbf{41,} 203001 (2008).
\end{thebibliography}
\end{document}